\documentclass[a4paper,fleqn]{cas-sc}

\usepackage[numbers,sort&compress]{natbib}

\def\tsc#1{\csdef{#1}{\textsc{\lowercase{#1}}\xspace}}
\tsc{WGM}
\tsc{QE}
\usepackage{amsmath,amssymb}
\usepackage{booktabs}
\usepackage{graphicx}
\usepackage{xcolor}
\usepackage{url}
\usepackage{xspace}
\usepackage{capt-of}
\usepackage{placeins}
\usepackage{array}

\newcommand{\kwqft}{\texttt{kwqft}\xspace}
\newcommand{\suN}{\mathrm{SU}(N)}
\newcommand{\tr}{\operatorname{Tr}}
\renewcommand{\Re}{\operatorname{Re}}

\begin{document}
\let\WriteBookmarks\relax
\def\floatpagepagefraction{0.8}
\def\textpagefraction{0.15}

\shorttitle{Performance portable $\mathrm{SU}(N)$ lattice gauge theory simulation with Kokkos}
\shortauthors{W. Sun}

\title[mode=title]{Performance Portable $\mathrm{SU}(N)$ Lattice Gauge Theory Simulation with Kokkos}

\author[1]{Wei Sun}[orcid=0000-0002-6350-8966]
%\cormark[1]
\ead{sunwei@ihep.ac.cn}
\credit{Conceptualization, Methodology, Software, Validation, Investigation,
  Data curation, Visualization, Writing -- original draft, Writing -- review \& editing}

\affiliation[1]{organization={Institute of High Energy Physics, Chinese Academy of Sciences},
  addressline={Yuquan Road 19B},
  city={Beijing},
  postcode={100049},
  country={People's Republic of China}}

\cortext[1]{Corresponding author}

\begin{abstract}
  The increasing diversity of high performance computing systems makes
  separate, architecture specific implementations of lattice gauge theory
  algorithms costly to maintain. We present \texttt{kwqft}, a performance
  portable Kokkos implementation of Wilson pure gauge Monte Carlo simulation for
  $\mathrm{SU}(N)$ Yang-Mills theory in an arbitrary number of space-time
  dimensions. The gauge group order $N$ and the dimension $D$ are compile
  time parameters. A single source targets the Serial, OpenMP, CUDA, HIP,
  and SYCL execution spaces, with MPI halo exchange overlapped with interior
  updates. The implementation reproduces the exact two-dimensional plaquette
  and published three and four dimensional values for gauge groups up to
  $\mathrm{SU}(17)$. On an NVIDIA A100 the Kokkos CUDA backend is competitive
  with a native CUDA code, SIMD acceleration improves the OpenMP path on
  Armv9 processors, and a large scale speedup is demonstrated for an
  $\mathrm{SU}(4)$ lattice on the LineShine supercomputer, currently ranked first on the TOP500
  list.
\end{abstract}

% \begin{highlights}
% \item One Kokkos source targets Serial, OpenMP, CUDA, HIP, and SYCL backends
% \item Host builds batch lattice sites with NEON, SVE, AVX2, or AVX-512 SIMD lanes
% \item Kokkos CUDA reaches 81\%--99\% of a native CUDA code on an NVIDIA A100
% \item Plaquettes verified against exact 2D results and published 3D/4D data
% \item MPI+OpenMP SU(4) strong scaling to 512 LineShine LX2 nodes
% \end{highlights}


\maketitle

\section{Introduction}\label{sec:introduction}

Lattice gauge theory turns the Euclidean path integral of a gauge field into
a finite dimensional statistical problem that can be evaluated by
Markov chain Monte Carlo methods \cite{Metropolis1953,GattringerLang2010}.
Wilson's formulation \cite{Wilson1974} made the theory gauge invariant
at nonzero lattice spacing and remains the basis of both production
calculations and methodological studies. Although present day lattice QCD
campaigns are dominated by dynamical fermion simulations, pure gauge
configuration generation and local gauge field measurements remain important in
several contexts: they are the first stage of quenched and partially
quenched studies, a testbed for algorithm development and code verification,
and the ideal place for studying $\suN$ Yang-Mills theory as a
function of the number of colors $N$ and of the space-time dimension $D$.
Large-$N$ studies of the glueball spectrum, string tensions, topology, and the
deconfinement transition in $2+1$ and $3+1$ dimensions
\cite{Teper:1998te,LuciniTeperWenger2004,LuciniPanero2013,Athenodorou:2016ebg,Athenodorou:2021qvs}
are entirely based on pure gauge ensembles, and they require simulation
software in which neither $N$ nor $D$ is hard coded.

The hardware used for lattice gauge theory simulations has become heterogeneous. Multicore
x86 and Arm processors with wide vector units coexist with accelerators from
several vendors, each with a different programming interface and preferred
memory access pattern. Maintaining a CUDA version, an OpenMP version, and
additional HIP or SYCL ports of the same algorithm duplicates implementation
and verification effort. Performance portability aims to make one
implementation useful across a declared set of platforms without forfeiting
an unacceptable fraction of the performance available on each platform
\cite{Pennycook2017}. This is a much stronger requirement than source
portability, because successful compilation alone does not guarantee efficient
execution.

Graphics processing units were adopted early for lattice gauge theory
\cite{Egri2007}, and mature CUDA libraries such as QUDA \cite{Clark2010QUDA}
have set the performance standard for fermion solvers. 
The pioneering work of Cardoso and Bicudo \cite{CardosoBicudo2013} used CUDA directly to generate four-dimensional $\suN$
pure gauge configurations and demonstrated that pseudo-heatbath, overrelaxation, and plaquette measurement
map efficiently to GPUs. Its architecture-specific CUDA source and fixed four-dimensional formulation, however, do not address portability across accelerator vendors and CPU architectures, nor general space-time dimensions, which are central design goals of this work.

Several portable programming models have been applied to lattice QCD. The
Grid library \cite{Boyle:2015tjk} achieves SIMD (Single Instruction Multiple Data) efficiency on CPUs by
distributing the lattice over ``virtual nodes'' so that one vector lane
holds one lattice site, an approach that also underlies the host SIMD path
described in this paper. Bonati et al.\ \cite{Bonati2017OpenACC,Bonati2018OpenACC}
used OpenACC directives to run a staggered fermion Monte Carlo code on
GPUs and multicore CPUs. 
On the other hand, Kokkos \cite{Edwards2014,Trott2021,Trott2022Kokkos3}
provides C++ abstractions for execution policies, multidimensional data,
reductions, and memory spaces and has been adopted by scientific applications
that must target both CPUs and GPUs. Jo{\'o} et al.\ \cite{Jo2019PerformancePO}
developed a Wilson Dslash mini-application with Kokkos and SYCL and evaluated its
single process performance on Intel Xeon, Xeon Phi, and NVIDIA GPU systems.
Their study showed that portable abstractions can deliver competitive
single device stencil performance, while also identifying cases where
explicit SIMD types, data layout choices, and architecture aware launch
policies were still needed. More recently, Schlepphorst and Krieg
\cite{Schlepphorst:2023apr} benchmarked a Kokkos and MPI lattice QCD kernel on
AMD EPYC, AMD MI250, Fujitsu A64FX, NVIDIA A100, and NVIDIA H100 systems,
demonstrating multi-node and multi-architecture feasibility. Production
packages such as SIMULATeQCD \cite{SIMULATeQCD2024} provide a complementary
reference for the organization, validation, and performance reporting of a
mature multi-GPU lattice QCD framework.

These studies establish the relevance of portable programming models to lattice
QCD, but they focus primarily on fermion operators or on the SU(3) gauge group.
The question addressed here is whether a complete pure gauge
update and measurement path can be
expressed once in Kokkos for general $N$ and $D$ while retaining competitive
performance from one processor to a large scale distributed memory supercomputer.
We present \kwqft\ (Kokkos Ken Wilson Quantum Field Theory), a compact Wilson pure gauge simulation program that
answers this question affirmatively. The contributions of this work are the
following.
\begin{enumerate}
\item A dimension-generic, compile-time parameterized $\suN$ implementation
      of the anisotropic Wilson gauge action, including pseudo-heatbath and
      overrelaxation updates and the standard observables, together with a
      distributed memory layer that performs halo exchanges needed by the Wilson
      staple and overlaps communication with interior updates.
\item A mapping of the data and parallel work to Kokkos views, parallel
      patterns, reductions, and random number pools without backend specific
      branches in the physics kernels.
\item A cross-site host SIMD path built on Kokkos \texttt{simd}
      with ISA-specialized de-interleaving loads and stores for ARM NEON/SVE (Scalable Vector Extension)
      and x86 AVX2/AVX-512, which leaves the per-site device kernels unchanged.
\item Verification of the implementation against the exact solution of
      two-dimensional $\suN$ lattice gauge theory for $N$ up to 15 and against
      published three and four dimensional plaquette values for $N$ up to 17.
\item A performance study comprising a comparison with a native CUDA code on
      an NVIDIA A100, the SIMD speedup on Armv9 processors, MPI+OpenMP strong
      scaling up to 512 nodes of the LineShine supercomputer, and the
      $N$ dependence of the trajectory time on six CPU and GPU platforms.
\end{enumerate}

The remainder of the paper is organized as follows.
Section~\ref{sec:physics} recalls the lattice formulation and the update
algorithms. Section~\ref{sec:software} describes the software design and its
mapping to Kokkos, including the distributed memory layer and the host SIMD
path. Section~\ref{sec:verification} reports the physics verification.
Sections~\ref{sec:method} and \ref{sec:results} present the benchmark
methodology and results, and Section~\ref{sec:conclusions} concludes.

\section{Lattice formulation and update algorithm}\label{sec:physics}

\subsection{Wilson pure gauge action}

Consider a $D$-dimensional hypercubic lattice with periodic boundary
conditions, extents $L_0,\ldots,L_{D-1}$, and volume $V=\prod_\mu L_\mu$. A
link variable $U_\mu(x)\in\suN$ connects site $x$ to $x+\hat\mu$. For an
isotropic lattice the Wilson action is \cite{Wilson1974}
\begin{equation}
 S_W[U]=\beta\sum_x\sum_{\mu<\nu}
 \left(1-\frac{1}{N}\Re\tr U_{\mu\nu}(x)\right),
 \label{eq:wilson}
\end{equation}
where $\beta=2N/g_0^2$ is the inverse bare coupling and
\begin{equation}
 U_{\mu\nu}(x)=U_\mu(x)\,U_\nu(x+\hat\mu)\,
 U_\mu^\dagger(x+\hat\nu)\,U_\nu^\dagger(x)
\end{equation}
is the oriented plaquette. 
For anisotropic gauge action with a bare anisotropy
parameter $\xi_0$ in direction $D-1$ designated as time, spatial--spatial
plaquettes receive the weight $\beta/\xi_0$ and plaquettes containing the
temporal direction receive $\beta\xi_0$. We focus on the isotropic case with $\xi_0=1$ in this paper.

For a selected link $U_\mu(x)$, the terms of Eq.~\eqref{eq:wilson} that
depend on it can be collected into the staple
\begin{equation}
 \Sigma_\mu(x)=\sum_{\nu\ne\mu} w_{\mu\nu}
 \Big[
  U_\nu(x+\hat\mu)\,U_\mu^\dagger(x+\hat\nu)\,U_\nu^\dagger(x)
  +U_\nu^\dagger(x+\hat\mu-\hat\nu)\,
  U_\mu^\dagger(x-\hat\nu)\,U_\nu(x-\hat\nu)
 \Big],
 \label{eq:staple}
\end{equation}
where $w_{\mu\nu}\in\{1/\xi_0,\xi_0\}$ is the anisotropy weight, so that
the local action is $S_{\rm loc}=-(\beta/N)\Re\tr[U_\mu(x)\Sigma_\mu(x)]$ up
to a constant. The staple involves $2(D-1)$ products of three links and
reads $6(D-1)$ distinct links: $U_\nu(x)$ at the updated site,
$U_\nu(x+\hat\mu)$ at the forward neighbor, and $4(D-1)$ link variables
on the neighboring sites $x\pm\hat\nu$ and $x+\hat\mu-\hat\nu$. It is the
only stencil operation in the update algorithms.

\subsection{Cabibbo--Marinari pseudo-heatbath}

A heatbath update replaces $U_\mu(x)$ by a new element drawn from the
conditional distribution $\propto\exp[(\beta/N)\Re\tr(U\Sigma)]$. For
SU(2) this distribution can be sampled directly \cite{Creutz1980}. Writing
$U\Sigma$ in quaternion form $a_0+i\,\boldsymbol a\cdot\boldsymbol\sigma$,
where $a_0^2+|\boldsymbol a|^2=1$ and $\boldsymbol\sigma$ denotes the vector of
Pauli matrices, the only nontrivial step is to draw $a_0\in[-1,1]$ from the
probability density $\propto\sqrt{1-a_0^2}\,e^{k a_0}$ with
$k=\beta\sqrt{\det\Sigma}$.
To keep the acceptance rate of this draw high, we use the algorithm of Kennedy
and Pendleton \cite{KennedyPendleton1985} in the large-$k$ regime and Creutz's
algorithm \cite{Creutz1980} for small $k$.

For $N>2$ a direct sampler is not available, and the pseudo-heatbath algorithm of Cabibbo
and Marinari \cite{CabibboMarinari1982} is used instead. A sequence of
SU(2) heatbath updates is applied in the $N(N-1)/2$ $\mathrm{SU}(2)_{pq}$ subgroups
that act on the color pairs $(p,q)$, $p<q$, and each SU(2)
matrix is embedded into the identity of $\suN$. 
For $N=3$ the three $2\times2$ blocks of
$U\Sigma$ needed by the successive subgroups are computed from the relevant
rows of $U$ and columns of $\Sigma$ without forming the full product. For
general $N$ the product $M=U\Sigma$ is formed once and both $U$ and $M$ are
then updated in place by each embedded SU(2) matrix, which requires
$\mathcal{O}(N)$ operations per subgroup instead of an $\mathcal{O}(N^3)$
matrix product. The total cost of one link update is therefore dominated by
the $\mathcal{O}(N^3)$ staple construction and the single
$\mathcal{O}(N^3)$ product $U\Sigma$, with an $\mathcal{O}(N^3)$ subgroup
term $N(N-1)/2\times\mathcal{O}(N)$ of comparable size.

\subsection{Overrelaxation}

The heatbath is complemented by the microcanonical overrelaxation update
\cite{BrownWoch1987,deForcrand:2005xr}, which moves the link as far as
possible on the surface of constant local action and thereby reduces
autocorrelation times at essentially no cost in acceptance.
The update is deterministic and does not change the action, so it cannot
be used alone as an update algorithm.
In each SU(2) subgroup the normalized projection of the $2\times2$ block of
$M=U\Sigma$ is formed, and the link is reflected within that subgroup. 
One trajectory in \kwqft\ consists of $n_{\rm hb}$ pseudo-heatbath sweeps followed
by $n_{\rm ovr}$ overrelaxation sweeps. All timing measurements in
Section~\ref{sec:results} use $n_{\rm hb}=1$ and $n_{\rm ovr}=1$ unless
stated otherwise, and the native CUDA comparison uses heatbath sweeps only.

\subsection{Checkerboard parallelism and reunitarization}

The staple of $U_\mu(x)$ depends only on links located on sites of the
opposite parity $(-1)^{\sum_\nu x_\nu}$ and on links in directions
$\nu\ne\mu$ of the same parity. The lattice is therefore split into even and
odd checkerboards, and for a fixed direction $\mu$ all $V/2$ links of one
parity can be updated concurrently. A complete sweep visits both parities and
all $D$ directions, giving $2D$ dependent kernel launches per sweep. The
update of each link is an independent work item, and the assignment of work
items to threads or SIMD lanes does not affect the Markov chain beyond the
usual rounding differences discussed in Section~\ref{sec:simd}.
After each trajectory the rows of every link matrix are orthonormalized with
a Gram--Schmidt procedure to remove the accumulated floating-point
drift away from the group manifold.

\subsection{Observables}

The average plaquette, normalized to unity for a trivial configuration, is
\begin{equation}
 P=\frac{2}{N\,D(D-1)\,V}
 \sum_x\sum_{\mu<\nu}\Re\tr U_{\mu\nu}(x).
 \label{eq:plaquette}
\end{equation}
The implementation separately accumulates spatial--spatial and
spatial--temporal plaquettes, which is useful for anisotropic lattices and
for monitoring the microcanonical character of the overrelaxation step. With
direction $D-1$ designated as time, the Polyakov loop is
\begin{equation}
 L_P=\frac{1}{N V_s}\sum_{\boldsymbol{x}}
 \tr\prod_{t=0}^{L_{D-1}-1} U_{D-1}(\boldsymbol{x},t),
 \label{eq:polyakov}
\end{equation}
where $V_s=V/L_{D-1}$ is the spatial volume. The product along the temporal
direction is a serial chain of $L_{D-1}$ matrix products for one spatial
site, while different spatial sites are reduced in parallel.

\section{Software design and Kokkos mapping}\label{sec:software}

\subsection{Scope and build configuration}

\kwqft\ is written in C++20 and built with CMake. Kokkos is
its only mandatory dependency. The build can use an installed Kokkos, a
local source tree, or Kokkos fetched from its public repository. MPI is
an optional dependency. The gauge group order $N$ (\texttt{KWQFT\_NCOLORS})
and the space-time dimension $D$ (\texttt{KWQFT\_NDIMS}) are compile time definitions
with SU(3) in four dimensions as the default. Making $N$ and $D$ compile time
constants allows all color loops and coordinate loops to be unrolled or
vectorized by the compiler and keeps small matrices in registers, while the
cost is that the software must be rebuilt once per $(N,D)$ pair, which takes
minutes with CPU and device compilers for large $N$. The build produces a library,
the \texttt{heatbath} executable, and a unit test executable.

The execution space is selected at configure time by one of the
options \texttt{KWQFT\_ENABLE\_OPENMP}, \texttt{KWQFT\_ENABLE\_CUDA},
\texttt{KWQFT\_ENABLE\_HIP}, or \texttt{KWQFT\_ENABLE\_SYCL}. When none is
given, the Serial backend is used. Accelerator and CPU architecture flags are
passed through the standard \texttt{Kokkos\_ARCH\_*} options
(e.g.\ \texttt{AMPERE80}, \texttt{AMD\_GFX90A}, or \texttt{NATIVE}), while the
application-level source code is left unchanged. Host OpenMP and Serial builds enable
cross-site SIMD vectorization by default (Section~\ref{sec:simd}). For
device backends the option \texttt{KWQFT\_MPI\_DEVICE\_AWARE} passes device
pointers directly to MPI; otherwise, halo buffers are staged through pinned
host memory. On HIP, where the runtime launches 1024-thread blocks
regardless of register usage, a \texttt{Kokkos::LaunchBounds} limit of 256
threads per block is applied to the register-heavy $\suN$ kernels. On CUDA the
occupancy calculator sizes the blocks and no bound is imposed. CUDA, HIP, and
OpenMP builds are represented in the performance evidence below. 
The SYCL build path was configured but not benchmarked, 
as no Intel discrete GPU was available for this study.

\subsection{Gauge field representation}

A $D$-dimensional $\suN$ lattice gauge field contains $DV$ link matrices, with $N^2$ complex elements each. The
field is stored in a structure-of-arrays (SOA) layout, and for a fixed matrix
element $(i,j)$, the values of successive links are contiguous in memory,
\begin{equation}
 \texttt{ptr}[\,\ell + (i N + j)\,S\,],\qquad
 \ell = x_{\rm eo} + \mu V,\quad S = DV,
 \label{eq:soa}
\end{equation}
where $x_{\rm eo}\in[0,V)$ is the even--odd site index (even sites first) and
$\mu$ the space-time direction. Each complex number is stored as an interleaved
(real, imaginary) pair. This layout has two properties that are exploited on
different platforms. On GPUs, neighboring threads processing neighboring
sites read neighboring addresses for each matrix element, so global memory
transactions are coalesced. On CPUs, the same matrix element from $W$
consecutive sites forms one contiguous block of $2W$ reals, which is exactly
the operand of a de-interleaving vector load (Section~\ref{sec:simd}).

Two storage formats are implemented. The full format stores all $N^2$
complex elements. For SU(3) the two-row format stores only the first two rows
($12$ reals) and reconstructs the third row as the complex conjugate of the
cross product of the first two, halving the memory traffic of the staple
\cite{Clark2010QUDA,CardosoBicudo2013}.
Allocation and memory space selection are managed by
\texttt{Kokkos::View} in the default memory space of the selected execution
space, and host mirrors are created only where host access is needed (e.g. I/O and
halo staging). The field type therefore contains no execution space details.

\subsection{Indexing, neighbor access, and genericity}

Coordinate and neighbor operations are templated over $D$, and matrix
operations over $N$. The even--odd index is decoded to coordinates with a
small number of integer divisions, and neighbors are obtained by
incrementing one coordinate with periodic wrap-around. Rather than
precomputing neighbor tables, the staple kernel uses lazily evaluated
``shift views'': a shifted field $U_\nu(x+\hat\mu)$ is a lightweight object
that resolves the source address on access, so no additional memory is
touched and the compiler can fuse the address arithmetic of the $2(D-1)$
staple legs. The design is inspired by the shift semantics of QDP++/Chroma
\cite{Chroma2005}. In MPI builds, the resolution returns either a local
address or an address in the received ghost buffer.
The checkerboard mapping requires an even extent in the first lattice
direction and, in MPI builds, an even local extent in every decomposed
direction so that parity is globally consistent across ranks. These
constraints are checked at start-up.

\subsection{Node level mapping to Kokkos}

Table~\ref{tab:kernels} summarizes the decomposition of the major
operations. A pseudo-heatbath or overrelaxation kernel is a one-dimensional
\texttt{Kokkos::RangePolicy} over the $V/2$ sites of one parity, and the parity
and direction loops remain on the host and issue $2D$ dependent launches per
sweep. Each work item loads the neighboring links, constructs the staple,
updates all embedded SU(2) subgroups, and writes back one link. Because a
work item reads only links of the other parity (or of the same parity in
different directions), neither atomics nor synchronization within a kernel
is needed. Reunitarization is a \texttt{parallel\_for} over all $DV$ links.
Plaquette and Polyakov loop measurements use \texttt{parallel\_reduce}
with scalar or small array reducers.
\begin{table}[t]
\caption{Mapping of the major operations to Kokkos abstractions.}
\label{tab:kernels}
\begin{tabular*}{\tblwidth}{@{\extracolsep{\fill}}lll@{}}
\toprule
Operation & Parallel domain & Kokkos mechanism \\
\midrule
Heatbath update & $V/2$ sites per parity and direction & \texttt{parallel\_for} \\
Overrelaxation & $V/2$ sites per parity and direction & \texttt{parallel\_for} \\
Reunitarization & $DV$ links & \texttt{parallel\_for} \\
Plaquette & $V$ sites & \texttt{parallel\_reduce} \\
Polyakov loop & $V_s$ spatial sites & \texttt{parallel\_reduce} \\
Random sampling & one stream per checkerboard site & \texttt{Random\_XorShift64\_Pool} \\
Halo packing & ghost links per (direction, parity) block & \texttt{parallel\_for} + \texttt{deep\_copy} \\
\bottomrule
\end{tabular*}
\end{table}

Random numbers are provided by
\texttt{Kokkos::Random\_XorShift64\_Pool} \cite{Marsaglia2003}. The pool is
sized to the number of checkerboard sites $V/2$, and the work item for site
$x$ acquires the generator state with index $x_{\rm eo}$ rather than an
arbitrary free state. Each site therefore owns one random stream that is
consumed in the same order on every backend and with every thread count,
which makes the random number sequence itself independent of the execution
space. Identical streams do not
imply bitwise-identical trajectories across backends, because different
compilers contract floating-point operations differently and the
Kennedy--Pendleton sampler contains data dependent rejection loops.
Agreement between backends is therefore established statistically
(Section~\ref{sec:verification}).

Timings are taken with \texttt{Kokkos::Timer} after a \texttt{Kokkos::fence()}.
The code reports for each kernel the elapsed time together with
floating-point operation count and memory access count. A complex multiplication is
counted as six real operations and a complex addition as two, so one
$\mathrm{SU}(N)$ product costs $6N^3+2N^2(N-1)$ operations. The staple on one
link is $4(D-1)$ such products, two on each of the $2(D-1)$ legs. The
Cabibbo--Marinari arithmetic on that link does not depend on $D$ and is
counted from the embedded $\mathrm{SU}(2)$ operations. Each work item moves
$6(D-1)+2$ link matrices and touches two random-number states. A sweep sums
over both parities and all $D$ directions. These figures are diagnostics for
comparing runs of the same code, and they are not used to claim absolute
efficiency with respect to hardware peaks.

\subsection{Distributed memory decomposition}\label{sec:mpi}

For distributed memory execution the global lattice is partitioned over a
$D$-dimensional Cartesian process grid $p_0\times\cdots\times p_{D-1}$
supplied on the command line. The product of the grid extents must equal the number of MPI
ranks, and every global extent must be divisible by the corresponding grid
extent. Each rank owns one hyper-rectangular subdomain, and Kokkos executes
the local update and measurement kernels on the rank's CPU cores or
accelerator, so that MPI+OpenMP, MPI+CUDA, and MPI+HIP configurations use
identical kernels.

Ghost buffers are
organized as one contiguous block per (parity, $\mu$) pair, so that one
block is one MPI message. A single halo object is shared by all kernels
operating on a gauge field and tracks the validity of each block, and a
checkerboard update marks the block of the updated direction and parity
stale, and a consumer refreshes only stale blocks. Consequently heatbath,
overrelaxation, plaquette, and reunitarization never re-send data that
another kernel has already communicated.

Communication is overlapped with computation. For every (parity, $\mu$)
step the sites whose staple reads a ghost (``boundary sites'') are updated
first. The corresponding block is packed and sent with non-blocking MPI
while the remaining ``interior sites'' are updated, and the exchange is then
completed before the next dependent launch. Boundary and interior site lists
are built once per lattice geometry. With device backends the pack and
unpack operations are Kokkos kernels, and buffers are either handed to a
device-aware MPI or staged through pinned host memory.

\subsection{Host SIMD vectorization}\label{sec:simd}

In addition to thread-level parallelism through the OpenMP execution space,
host builds vectorize independent lattice sites with
\texttt{Kokkos::Experimental::simd}. The option
\texttt{KWQFT\_ENABLE\_HOST\_SIMD} is enabled by default for OpenMP and
Serial configurations (e.g. a pure MPI build) and is
rejected at configure time when CUDA, HIP, or SYCL is selected, so that
accelerator kernels keep one lattice site per device thread.

In the cross-site SIMD implementation, each SIMD lane corresponds to one lattice site of the
checkerboard update. A batch of $W=\texttt{simd::size()}$ sites is processed
together, where staple accumulation, link load and store, the dense $\suN$ algebra (a complex
general matrix product written with explicit \texttt{Kokkos::fma} to better exploit fused multiply-add instructions),
the heatbath, and the overrelaxation all run across lanes. 

The gauge field retains the interleaved complex SOA layout of
Eq.~\eqref{eq:soa}. When the $W$ sites of a batch address $W$ consecutive
links, each matrix element of the batch is $2W$ consecutive floating point numbers and is
moved with a single de-interleaving vector load or store specialized for the
active Kokkos SIMD ABI, i.e. ARM SVE (\texttt{svld2}/\texttt{svst2}), ARM NEON
(\texttt{vld2}/\texttt{vst2}), x86 AVX-512 (two source permutes), or x86 AVX2
(unpack/shuffle with a cross-lane permute). Kokkos provides no SSE ABI, so an
x86 build without AVX2 uses the scalar ABI ($W=1$) and reverts to the
per-site update. With \texttt{Kokkos\_ARCH\_NATIVE} the native width is
selected by $W=2$ (NEON), $4$ (AVX2), or $8$ (AVX-512 or SVE-512)
double precision sites.

\section{Verification}\label{sec:verification}

Correctness is established by comparing plaquette expectation values from
\kwqft\ with exact solutions for various $N$ in two dimensions, and with published Monte Carlo results at matched
lattice parameters in three and four dimensions. 
Each \kwqft\ ensemble consists of 20000 trajectories, which are generated from a cold start with one heatbath
sweep followed by one overrelaxation sweep per trajectory.
All of these ensembles, and the comparisons reported in this section, were produced with the Kokkos CUDA backend on a single NVIDIA A100 GPU.
The first $10000$ trajectories are discarded and the following $10000$ measurements are analyzed with a
blocked jackknife error estimator (block size 100) to account for autocorrelations. 
Agreement between different Kokkos backends
was checked in the same way on smaller lattices, and the ensembles generated by
different backends from the same seed agree within statistical
uncertainties, while they are not expected to be bitwise identical.

\subsection{Two dimensions: comparison with the exact solution}

Two-dimensional $\suN$ lattice gauge theory with the Wilson action is
exactly solvable \cite{GrossWitten1980,Drouffe:1983fv,Carlsson:2003wx}. In
axial gauge the plaquettes are independent, and up to exponentially small
finite-volume corrections the partition function factorizes into
single-plaquette integrals. With $x\equiv\beta/N$, the SU($N$) single-plaquette
integral is
\begin{equation}
  Z_N(x) = \sum_{q=-\infty}^{\infty} D_N^{(q)}(x),\qquad
  D_N^{(q)}(x) = \det_{0 \leq i, j < N} I_{q+i-j}(x),
\end{equation}
where $I_n(x)$ is the modified Bessel function of the first kind and the sum
over $q$ implements the unit-determinant constraint. Then the average plaquette is
\begin{equation}\label{eq:plaquette_analytic}
  \langle P_N \rangle = \frac{1}{N} \frac{d}{dx} \ln Z_N(x)
  = \frac{1}{N} \frac{\sum_q \big(D_N^{(q)}(x)\big)'}{\sum_q D_N^{(q)}(x)},
\end{equation}
and the derivative of each Toeplitz determinant is evaluated with
$I_n'(x) = \tfrac{1}{2}\big(I_{n-1}(x) + I_{n+1}(x)\big)$. For
$\mathrm{SU}(2)$ the expression reduces to
$\langle P_2 \rangle = I_2(\beta)/I_1(\beta)$. We truncate the sum at
$|q|\leq30$, which has converged to machine precision in the range of $\beta$
considered.

\begin{figure}[!htbp]
\centering
\includegraphics[width=0.74\linewidth]{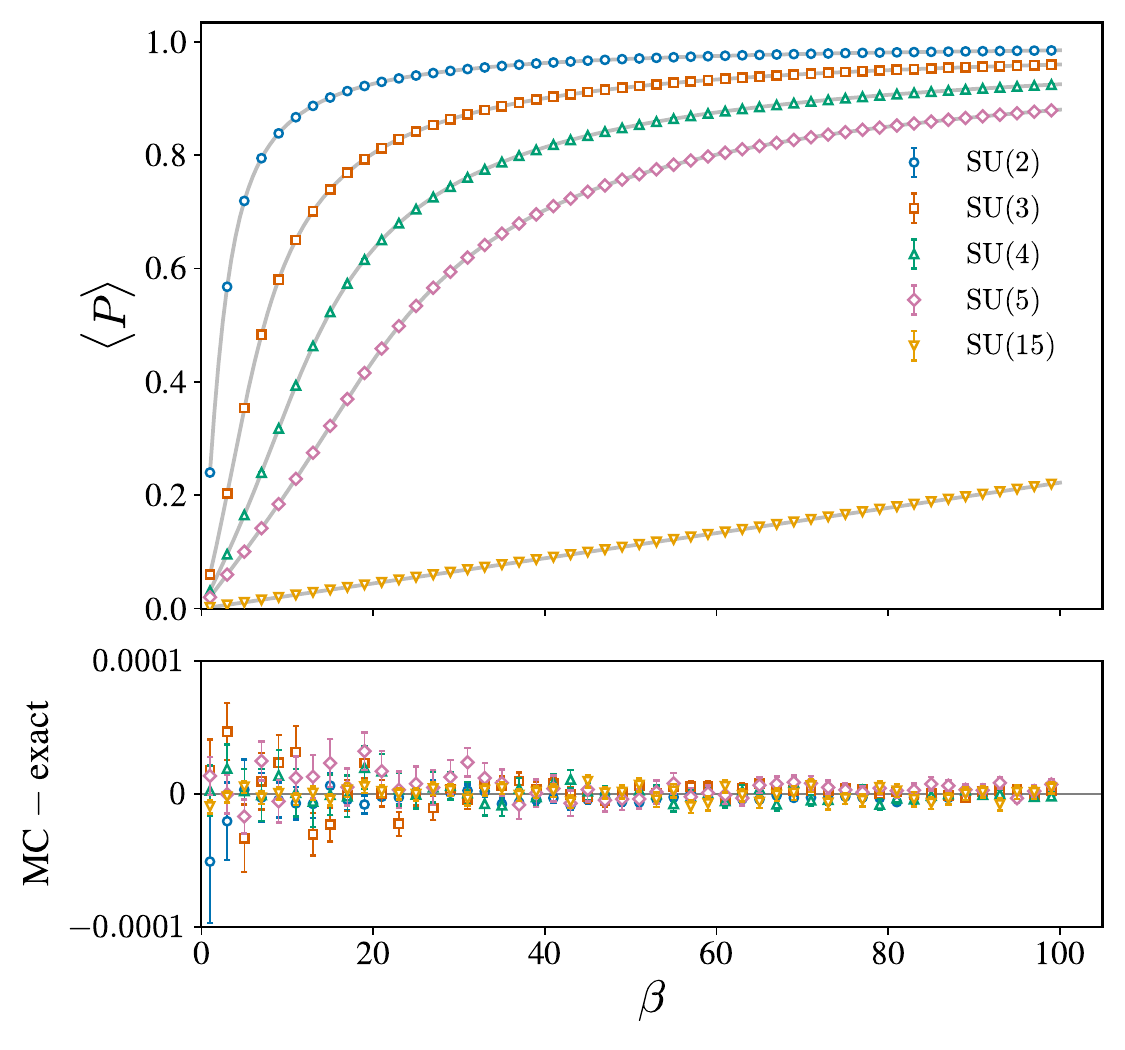}
\caption{Average plaquette $\langle P\rangle$ in two-dimensional
$\mathrm{SU}(N)$ Yang--Mills theory as a function of $\beta$. Curves are the
exact results from Eq.~\eqref{eq:plaquette_analytic}, markers are \kwqft\
Monte Carlo measurements on a $100\times100$ lattice ($20000$
trajectories, the last $10000$ analyzed with blocked jackknife errors). The
lower panel shows the residuals $\mathrm{MC}-\mathrm{exact}$.}
\label{fig:plaquette_2d}
\end{figure}

Figure~\ref{fig:plaquette_2d} compares Eq.~\eqref{eq:plaquette_analytic}
with \kwqft\ measurements on a $100\times100$ lattice for
$\mathrm{SU}(2)$ to $\mathrm{SU}(5)$ and $\mathrm{SU}(15)$ at $\beta$ values
between 1 and 100. The residuals in the lower panel show close agreement
between the exact solution and the Monte Carlo results from \kwqft\
over the entire range, including the
strong coupling region where $\langle P\rangle$ varies rapidly with $\beta$
and the weak coupling region where, for small $N$, the plaquette approaches
unity and the statistical errors become very small.

\subsection{Three and four dimensions: comparison with published data}

In three dimensions we compare with the high-precision plaquette values of
Hietanen and Kurkela \cite{Hietanen:2006rc}, obtained for $N=2$--$8$ on
lattices up to $320^3$ (Table~\ref{tab:plaquette_3d}). In four dimensions we
compare with the ensembles of Bali et al.\ \cite{Bali:2013kia}, which cover
$N=2$--$7$ and $N=17$ at couplings chosen to keep the string tension in
lattice units approximately constant (Table~\ref{tab:plaquette_4d}). In 21
of the 22 cases the \kwqft\ value agrees with the reference within $2.3$
combined standard deviations, and in 16 cases within one. For three parameter
sets the central values are exactly the same: $N=2$, $\beta=16$ on
$320^3$ in Table~\ref{tab:plaquette_3d}, and $N=5$, $\beta=17.535$ together
with $N=7$, $\beta=34.8343$ in Table~\ref{tab:plaquette_4d}. The published
results do not specify the Monte Carlo algorithm or the hardware platform
that produced them, and the \kwqft\ ensembles were not tuned to either, so
these identical digits are obtained without reproducing those choices. The exception is
$N=8$, $\beta=100$ in three dimensions, where the two values differ by
$3.2\times10^{-6}$ (a relative difference of $4\times10^{-6}$), or about
$4.5$ combined standard deviations, both uncertainties being at the
$5\times10^{-7}$ level. At this precision, residual autocorrelations and the
thermalization cut of the two independent ensembles matter, and neither
uncertainty estimate can be verified from the published information.
We therefore do not regard this case as evidence of a systematic error. 
These comparisons exercise the $N=2$, $N=3$, and general-$N$ code
paths of the heatbath, the overrelaxation, and the plaquette in both odd and
even dimensions.

\begin{table}[t]
  \centering
  \caption{Average plaquette $\langle P\rangle$ in three-dimensional
  $\mathrm{SU}(N)$ Yang--Mills theory. Reference values are from
  Ref.~\cite{Hietanen:2006rc} (converted from $1-\langle P\rangle$). \kwqft\
  values use the last $10000$ trajectories analyzed with blocked jackknife errors.}
  \label{tab:plaquette_3d}
  \small
  \begin{tabular}{rrlll}
    \toprule
    $N$ & $\beta$ & Lattice & $\langle P\rangle$, Ref.~\cite{Hietanen:2006rc} & $\langle P\rangle$, \kwqft \\
    \midrule
    2  & 6   & $48^3$     & 0.8247839(16)  & 0.8247783(40)\\
    2  & 16  & $320^3$    & 0.93649802(5)  & 0.93649802(7)\\
    2  & 32  & $320^3$    & 0.96851172(2)  & 0.96851176(4)\\
    3  & 12  & $24^3$     & 0.7582875(8)   & 0.7582951(101)\\
    3  & 24  & $48^3$     & 0.8849014(11)  & 0.8849019(14)\\
    3  & 80  & $128^3$    & 0.96634760(6)  & 0.96634748(10)\\
    4  & 24  & $48^3$     & 0.7742299(7)   & 0.7742292(27)\\
    4  & 50  & $64^3$     & 0.8966907(4)   & 0.8966914(7)\\
    4  & 100 & $128^3$    & 0.94923340(4)  & 0.94923347(11)\\
    5  & 40  & $128^3$    & 0.7838764(5)   & 0.7838765(5) \\
    5  & 100 & $128^3$    & 0.91791074(8)  & 0.91791095(14)\\
    5  & 180 & $160^3$    & 0.95494414(5)  & 0.95494409(5)\\
    8  & 100 & $96^3$     & 0.7714494(5)   & 0.7714526(5)\\
    8  & 300 & $96^3$     & 0.92839623(11) & 0.92839588(12)\\
    8  & 460 & $128^3$    & 0.95368978(5)  & 0.95368987(5)\\
    \bottomrule
  \end{tabular}
\end{table}

\begin{table}[t]
  \centering
  \caption{Average plaquette $\langle P\rangle$ in four-dimensional
  $\mathrm{SU}(N)$ Yang--Mills theory. Reference values are from Table~5 of
  Ref.~\cite{Bali:2013kia}. \kwqft\ values use the last $10000$ trajectories analyzed with blocked jackknife errors.}
  \label{tab:plaquette_4d}
  \small
  \begin{tabular}{rrlll}
    \toprule
    $N$ & $\beta$ & Lattice & $\langle P\rangle$, Ref.~\cite{Bali:2013kia} & $\langle P\rangle$, \kwqft \\
    \midrule
    2  & 2.4645  & $32^3\times 64$  & 0.644718(3)  & 0.644717(3)\\
    3  & 6.0175  & $32^3\times 64$  & 0.595595(3)  & 0.595596(2) \\
    4  & 11.0280 & $24^3\times 48$  & 0.578794(2)  & 0.578791(3)\\
    5  & 17.5350 & $24^3\times 48$  & 0.573069(2)  & 0.573069(2) \\
    6  & 25.4520 & $24^3\times 48$  & 0.568682(2)  & 0.568684(2)\\
    7  & 34.8343 & $24^3\times 48$  & 0.566504(2)  & 0.566504(2)\\
    17 & 208.45  & $12^3\times 24$  & 0.562729(4)  & 0.562733(3)\\
    \bottomrule
  \end{tabular}
\end{table}

\section{Performance methodology}\label{sec:method}

\subsection{Platforms}

Table~\ref{tab:platforms} lists the hardware and software configurations
used. They span two Arm server processors with NEON and SVE vector
units, one x86 processor with AVX-512, two NVIDIA GPUs from different
generations, and one Hygon Deep Computing Unit (DCU) programmed through the
HIP execution space. All measurements are in double precision, and all
kernels run in the default memory space of the selected backend (device
memory for GPUs and the DCU).

The distributed memory experiments were carried out on LineShine supercomputer~\cite{YutongLu2026}, 
the exascale HPC--AI system of the National
Supercomputing Center in Shenzhen. Its compute nodes are built from two LX2
CPUs, an Armv9 design with 304 cores per socket arranged as two dies
and eight clusters of 38 cores. A node therefore exposes 608 cores in 16
clusters. Every core has private 32~KB L1 instruction and data
caches, and the 38 cores of a cluster share a 28.5~MB L2 cache.
The memory system is two-tiered: each processor carries 32~GB of on-package
HBM with an aggregate peak bandwidth of 4~TB/s, backed by 256~GB of DDR
memory, and each die is divided into four HBM and four DDR NUMA domains,
one pair per cluster, so that a socket presents 16 NUMA domains and a node
32. HBM access is fast only from the cluster it is attached to, whereas DDR
bandwidth is shared more uniformly within a die. On the compute side the
cores implement SVE with a 512-bit vector length, which is the width used by
the SIMD path of Section~\ref{sec:simd}, as well as the Scalable Matrix
Extension (SME), which \kwqft\ has not yet exploited. The double precision peak of one
LX2 processor is 60.3~TFLOP/s. Nodes are connected by the LingQi network
with 1.6~Tb/s bandwidth per node, and MPI traffic in
our runs is carried by UCX over RoCE. In all LX2 runs we place one MPI rank
per cluster, bind its OpenMP threads to the cores of that cluster, and bind
its memory to the HBM domain of the same cluster, so that the gauge field
and halo buffers of a rank reside in locality-matched HBM. Runs described
as ``LineShine LX2 CPU'' below use eight ranks of 38 threads, i.e.\ half a
node.

\begin{table}[t]
\caption{Hardware and software platforms. LX2 core counts refer to one node.
``SIMD'' denotes the cross-site host SIMD path of Section~\ref{sec:simd}.}
\label{tab:platforms}
\small
\begin{tabular*}{\tblwidth}{@{\extracolsep{\fill}}
  >{\raggedright\arraybackslash}p{1.3cm}
  >{\raggedright\arraybackslash}p{3.3cm}
  >{\raggedright\arraybackslash}p{3.4cm}
  >{\raggedright\arraybackslash}p{2.7cm}
  >{\raggedright\arraybackslash}p{3.4cm}@{}}
\toprule
class & processor/accelerator & compute units & backend & compiler / runtime \\
\midrule
x86 CPU & Intel Xeon Gold 6240R &
  24 cores & OpenMP+SIMD (AVX-512) &
  GCC 11.4.1 \\
Arm CPU & Huawei Kunpeng 920 \cite{Xia2021Kunpeng} &
  48 cores & OpenMP+SIMD (NEON) &
  GCC 11.4.1 \\
Arm CPU & LineShine LX2 \cite{YutongLu2026} &
  $2\times304$ cores per node in 16 clusters of 38 cores & OpenMP+SIMD (SVE-512) &
  BiSheng 5.2.0; Hyper MPI 26.1 \\
NVIDIA GPU & A100 (40~GB, PCIe) &
  6912 CUDA cores / 108 SMs & CUDA &
  CUDA 12.4 \\
NVIDIA GPU & RTX 5060 Laptop GPU (8~GB) &
  3328 CUDA cores / 26 SMs & CUDA &
  CUDA 13.1 \\
Hygon DCU & BW1000 (64~GB) &
  80 compute units & HIP &
  DTK 25.04.4 \\
\bottomrule
\end{tabular*}
\end{table}

\subsection{Measurements and metrics}

The quantity reported throughout is the wall-clock \emph{trajectory time},
measured on rank 0 between fences around the update sweeps of one
trajectory, as in the native
CUDA reference~\cite{CardosoBicudo2013}. 
Each data point is the average over the last 50 or
100 trajectories of a run of 50 to 200 trajectories from a cold start, which
discards start-up effects such as cache-cold first launches and MPI
connection establishment. 
Four experiments are performed:
\begin{enumerate}
\item \emph{Kokkos versus native CUDA.} Four dimensional SU(4) at $\beta=6$ on isotropic
      $L^4$ lattices with $L=8$--$72$ on one A100, one heatbath sweep per
      trajectory in both codes. The metric is the ratio
      $T_{\rm CUDA}/T_{\rm Kokkos}$ of trajectory times at equal $L$, which
      we call the Kokkos efficiency. It quantifies the cost of the abstraction
      on one platform.
\item \emph{Host SIMD speedup.} $N=2$--$15$ on a $24^3\times64$ lattice at
      $\beta=6$ on one LX2 CPU (8 MPI ranks $\times$ 38 OpenMP
      threads, one rank per cluster),
      comparing the
      cross-site SIMD build with the scalar build of the same code.
\item \emph{MPI+OpenMP scaling on LX2.} SU(4) at $\beta=11.028$.
      Strong scaling keeps the global volume $128^3\times256$ fixed from
      16 to 512 LineShine LX2 nodes. Speedup and
      parallel efficiency are defined as
      \begin{equation}
       S(p)=\frac{T(p_0)}{T(p)}\,,\qquad
       E_{\rm strong}(p)=\frac{p_0}{p}\,S(p)\,,
       \label{eq:scaling}
      \end{equation}
      with $p_0=16$ nodes for the strong scaling baseline.
\item \emph{Trajectory time versus $N$ across platforms.} $N=2$--$15$ on a
      $24^3\times64$ lattice at $\beta=6$ on all six platforms, with the
      SIMD path active on the CPUs.
\end{enumerate}

\section{Performance results}\label{sec:results}

\subsection{Kokkos CUDA backend versus native CUDA}\label{sec:cuda}

\begin{figure}[!htbp]
\centering
\includegraphics[width=0.74\linewidth]{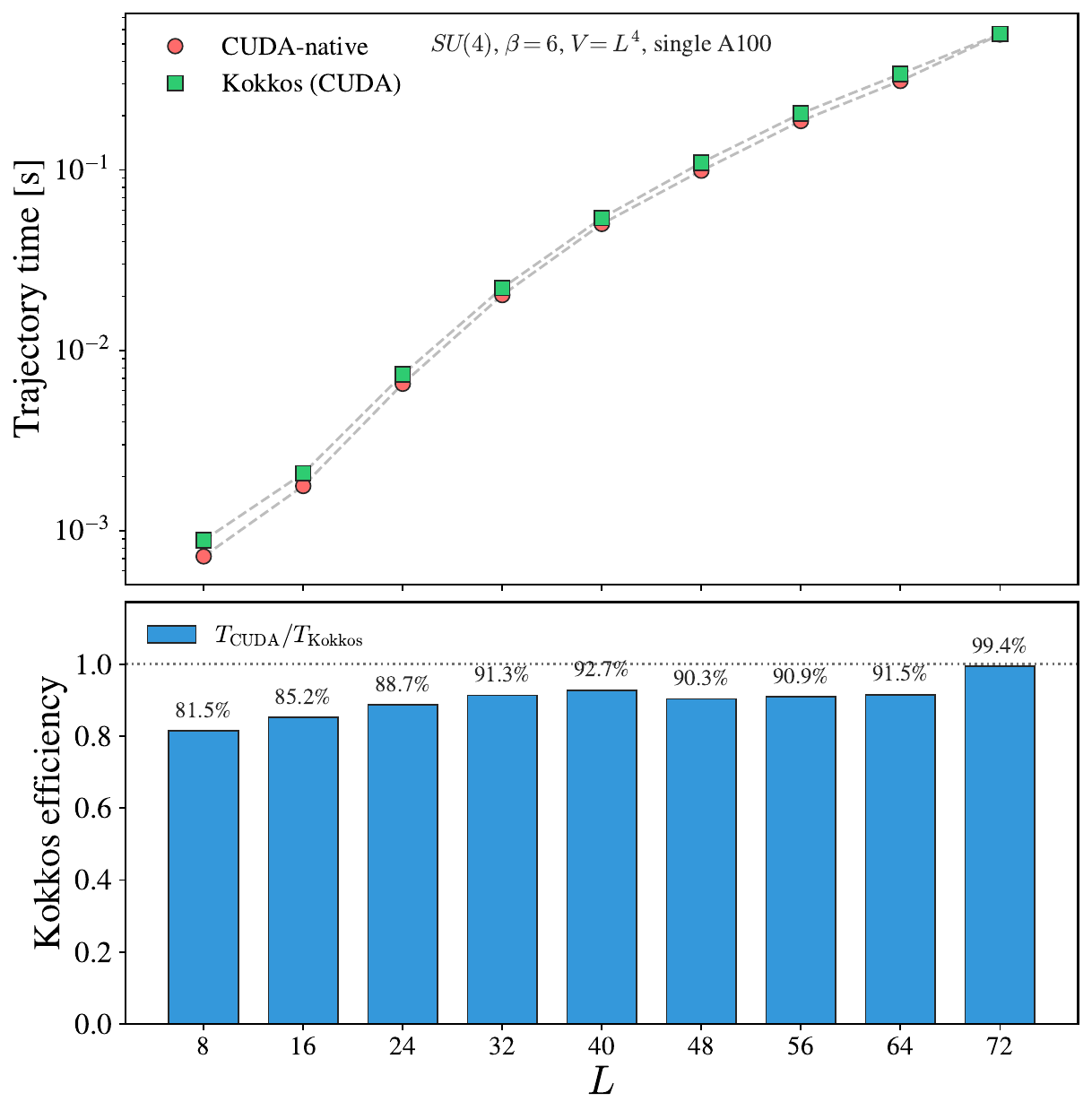}
\caption{Four dimensional SU(4), $\beta=6$ pseudo-heatbath on one NVIDIA A100. Top: average
trajectory time (one heatbath sweep) of the Kokkos CUDA backend and of the
native CUDA code versus $L$ for $L^4$ lattices. Bottom: Kokkos efficiency
$T_{\rm CUDA}/T_{\rm Kokkos}$.}
\label{fig:cuda}
\end{figure}

Figure~\ref{fig:cuda} compares the trajectory time of the Kokkos CUDA backend
with the native CUDA implementation \cite{CardosoBicudo2013} on
an NVIDIA A100 GPU. The two curves are nearly indistinguishable on the logarithmic
scale over more than two orders of magnitude in time. The lower panel
quantifies the ratio: the Kokkos efficiency rises from 81.5\% at $L=8$
through 85.2\% at $L=16$ and 88.7\% at $L=24$ to $90$--$93\%$ for
$L=32$--$64$ and 99.4\% at $L=72$. At the smallest volumes each launch
processes only $V/2=2048$ work items, fewer than the number of resident
warps on an A100, so fixed launch and synchronization costs dominate and
small differences in launch overhead between the Kokkos runtime and the
native CUDA code are exposed. Once the device is saturated the
remaining $7$--$10\%$ gap is attributable to differences in kernel
organization: the native code specializes the staple loop structure and
register usage for the fixed dimension, whereas the Kokkos kernel is
dimension-generic and relies on the compiler to resolve the lazy shift
addresses. The near-unity value at $L=72$ should not be over-interpreted.
It reflects that at this size both codes are bound by the same memory
traffic, and measurement variation at the percent level is comparable to
the difference. Overall, the abstraction does not impose a penalty for this kernel
on the A100 for lattices of practical size. \kwqft\ therefore meets the
performance portability requirement on this platform: one Kokkos source,
without a CUDA-specific kernel, delivers the throughput of the native
implementation.

\subsection{Host SIMD speedup on LineShine}\label{sec:simd_results}

Table~\ref{tab:simd_lx2} reports the trajectory time of the scalar and the
cross-site SIMD OpenMP builds on one LineShine LX2 CPU for $N=2$ to $15$. With
eight double precision lanes the SIMD build is $1.7$ to $2.1\times$ faster
for $N\leq10$ and $2.4$ to $3.5\times$ faster for $N=11$ to $15$. The speedup is
smaller than the nominal factor of eight because the compiler already vectorizes part of
the color loops within one site in the scalar build. The increase of the speedup with $N$
reflects the growing weight of the $\mathcal{O}(N^3)$ dense algebra, which
runs entirely on fused multiply add instructions across lanes, and the fact
that for large $N$ the scalar build suffers from the register and
instruction cache pressure.
These results show that the site-parallel structure
of the update is sufficient to fill wide vector units without changing
the color-loop structure that the GPU kernels rely on.

\begin{table}[t]
\centering
\caption{Trajectory time (one heatbath and one overrelaxation sweep,
$24^3\times64$ lattice, $\beta=6$) of the scalar and cross-site SIMD
(SVE-512, $W=8$) OpenMP builds on one LineShine LX2 CPU (8 MPI ranks
$\times$ 38 threads, one rank per cluster). Each entry is the mean over the last 50 trajectories
of a run.}
\label{tab:simd_lx2}
\small
\begin{tabular}{rrrr@{\hskip 2em}rrrr}
\toprule
$N$ & Scalar [s] & SIMD [s] & Speedup & $N$ & Scalar [s] & SIMD [s] & Speedup \\
\midrule
 2 & 0.073 & 0.043 & 1.70 &  9 & 1.280 & 0.687 & 1.86 \\
 3 & 0.131 & 0.071 & 1.84 & 10 & 1.774 & 0.893 & 1.99 \\
 4 & 0.248 & 0.119 & 2.08 & 11 & 2.750 & 1.135 & 2.42 \\
 5 & 0.380 & 0.179 & 2.12 & 12 & 3.957 & 1.428 & 2.77 \\
 6 & 0.522 & 0.264 & 1.98 & 13 & 5.918 & 1.732 & 3.42 \\
 7 & 0.637 & 0.374 & 1.70 & 14 & 7.427 & 2.120 & 3.50 \\
 8 & 0.902 & 0.512 & 1.76 & 15 & 8.512 & 2.562 & 3.32 \\
\bottomrule
\end{tabular}
\end{table}

\subsection{MPI+OpenMP scaling on LineShine LX2}\label{sec:lx2}

\begin{figure}[!htbp]
\centering
\includegraphics[width=0.74\linewidth]{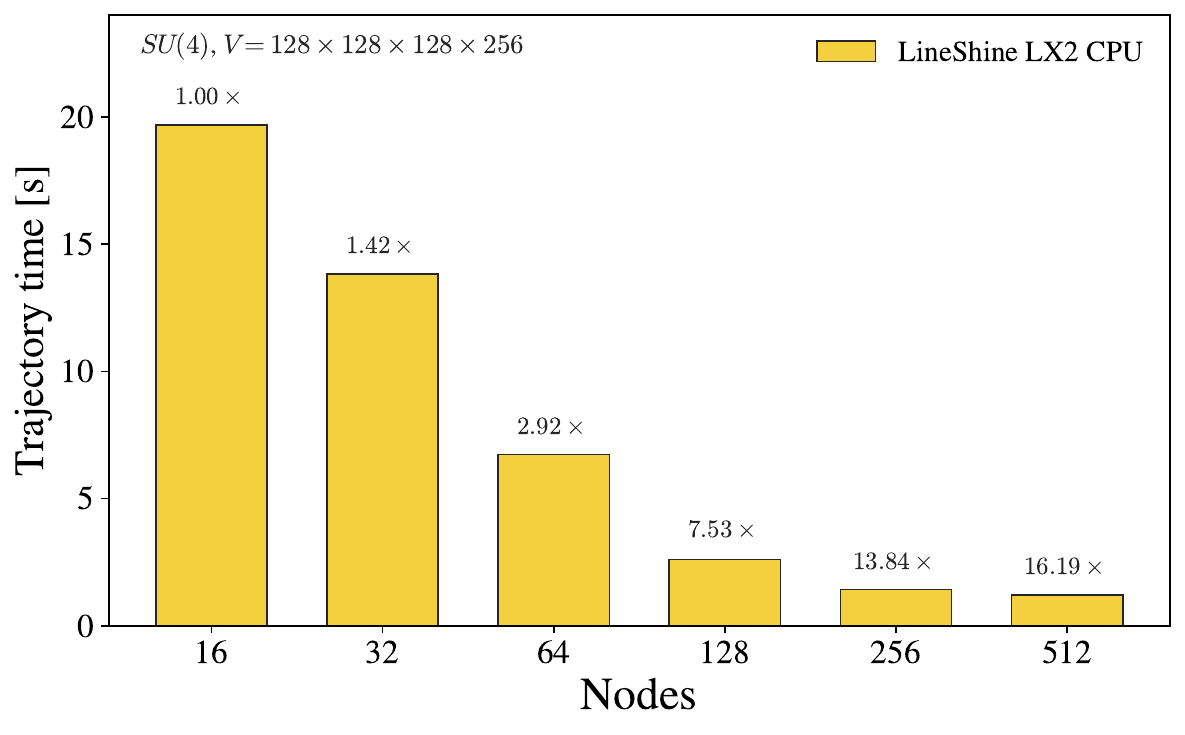}
\caption{MPI+OpenMP strong scaling of four-dimensional SU(4) lattice gauge
theory on LineShine LX2 CPUs.
Bars show the trajectory time (one heatbath and one overrelaxation sweep)
for the fixed volume $128^3\times256$ at $\beta=11.028$, averaged over the
last 100 of 200 trajectories, with labels giving the speedup relative to 16 nodes.}
\label{fig:lx2_su4}
\end{figure}

Figure~\ref{fig:lx2_su4} shows the speedup of the trajectory time for
the fixed $128^3\times256$ SU(4) lattice from 16 nodes to 512 nodes. The trajectory time falls from
19.5~s to 1.2~s, a speedup $S(512)=16.2$ for a $32\times$ increase in
resources, i.e.\ a strong scaling efficiency $E_{\rm strong}(512)=51\%$,
with $S(p)$ and $E_{\rm strong}(p)$ defined in Eq.~\eqref{eq:scaling}. 
The first doubling to 32 nodes yields only $1.42\times$,
while the subsequent doublings to 64 and 128 nodes yield $2.06\times$ and
$2.58\times$, i.e.\ super-linear gains, and the efficiency relative to
16 nodes recovers to 94\% at 128 nodes ($7.53\times$) before falling to 87\%
at 256 nodes ($13.8\times$) and 51\% at 512 nodes. The super-linear steps
coincide with the per-rank gauge field shrinking from 2.1~GB at 16 nodes to
0.27~GB at 128 nodes (1~KB per site for SU(4) in double precision), so that
a growing fraction of the field and of the halo buffers is served from the
28.5~MB cluster-level cache and from the cluster's own HBM domain, whose
bandwidth is strongly dependent on locality. At 512 nodes
each rank owns only a small sublattice. The ratio of
boundary to interior sites approaches unity, and the halo
exchanges per sweep can no longer be hidden behind the interior update, so that the
remaining time is dominated by message latency and by the synchronization
implied by the dependent launches.

\subsection{Trajectory time versus $N$ across architectures}\label{sec:trajN}

\begin{figure}[!htbp]
\centering
\includegraphics[width=0.84\linewidth]{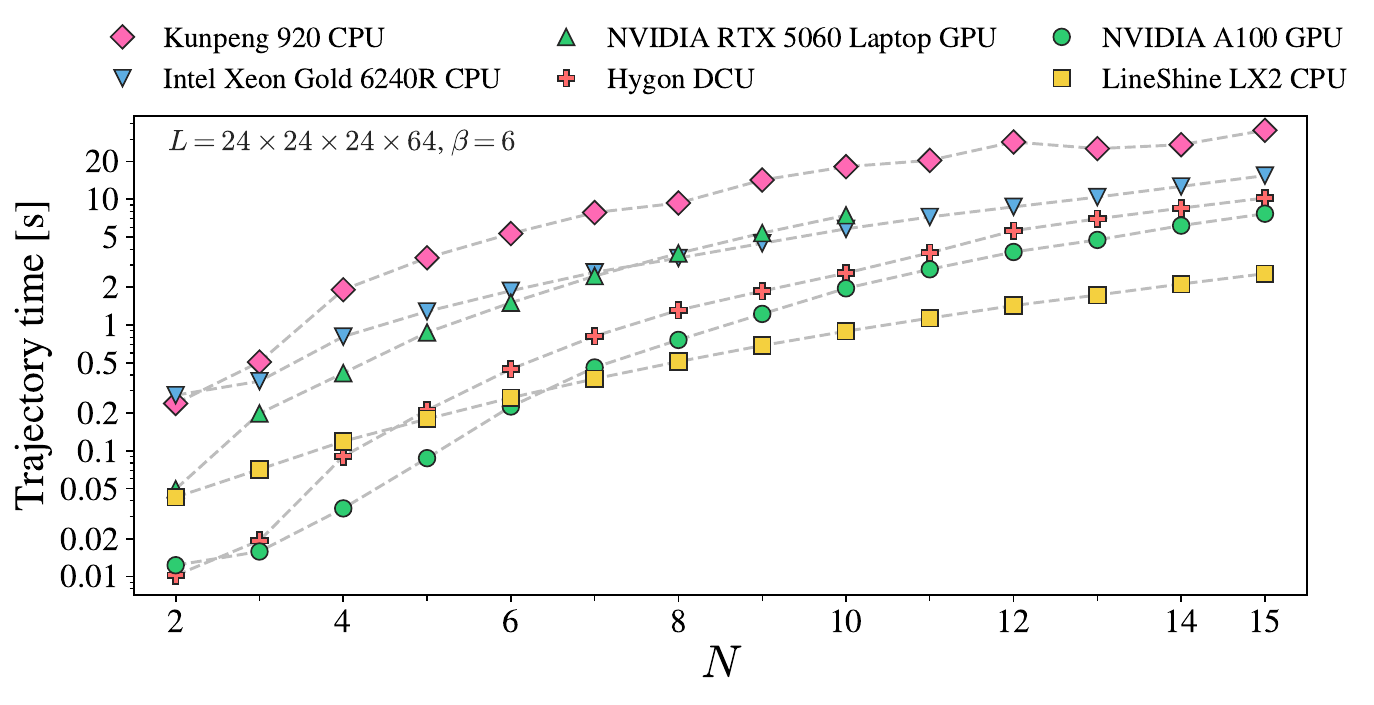}
\caption{Trajectory time (one heatbath and one overrelaxation sweep,
average over 50 trajectories) versus $N$ for four dimensional $\suN$ on a
$24^3\times64$ lattice at $\beta=6$: Kunpeng 920 (48 cores, NEON), Intel Xeon
Gold 6240R (24 cores, AVX-512), and one LineShine LX2 CPU (304 cores,
SVE-512) with the OpenMP+SIMD build; NVIDIA RTX 5060 Laptop GPU and NVIDIA
A100 with the CUDA build; Hygon DCU with the HIP build. The RTX 5060
series ends at $N=10$ because of its 8~GB memory.}
\label{fig:traj_N}
\end{figure}

Figure~\ref{fig:traj_N} compares the wall-clock trajectory time for
four-dimensional $\suN$ on a fixed $24^3\times64$ lattice at $\beta=6$ for
$N=2$--$15$ on the six platforms of Table~\ref{tab:platforms}. The same
application source is used on each platform. Only the CMake options, such as
architecture flags and backend configurations, change. On the CPU platforms
the host OpenMP build with cross-site SIMD is used, so that NEON, AVX-512,
or SVE-512 vector units are active according to the target, and the GPU and DCU
backends use the scalar per-site device kernels.

Three observations follow. First, the trajectory time grows with $N$ on
every platform with a common shape, approaching the $N^3$ behavior expected
from the operation count of Section~\ref{sec:software} for large $N$;
between $N=2$ and $N=15$ the time increases by a factor of about 60 on the
LX2 processor but by a factor of about 600 on the A100 and about 1000 on the
DCU. The accelerators show a much steeper rise at small $N$ than the CPUs
because their small-$N$ kernels are latency- and occupancy-bound (the SU(2)
trajectory takes about 10~ms on both the A100 and the DCU, and the
lattice has only $2.2\times10^5$ sites per parity). 
Second, for $N\geq7$ the ordering of the
platforms is stable: the LX2 CPU (304 SVE-512 cores) is fastest,
followed by the A100, the DCU, the RTX 5060, the Xeon Gold 6240R, and the
Kunpeng 920. The A100 is fastest for $N\leq5$, and the crossover with the LX2
CPU between $N=6$ and $N=7$ reflects the growing register pressure of the
per-thread $\suN$ kernels on the GPU, which reduces occupancy, whereas the
CPU SIMD path keeps its lanes fully utilized as $N$ grows. Third, the HIP
build on the DCU is within a factor of $1.3$--$2.6$ of the A100 for $N\geq4$
and faster than the RTX 5060 for all $N$, without any DCU-specific source
changes. This is direct evidence that the portability layer reaches a
non-NVIDIA accelerator ecosystem with useful performance.

\subsection{Interpretation}

The experiments cover three distinct aspects of performance portability.
The A100 comparison estimates the cost of replacing explicit CUDA
mechanisms by Kokkos abstractions on one accelerator: at most 19\% at the
smallest volume and below 10\% at practical volumes. The LX2 experiments
demonstrate that the same kernels, combined with the MPI layer, operate at
the scale of 512 nodes and that the host SIMD path delivers a $2$--$3.5\times$
gain on a 512-bit vector architecture that Kokkos supports through its SIMD
abstraction. Figure~\ref{fig:traj_N} shows that the $\suN$-parameterized code
path yields consistent $N$ scaling on two NVIDIA GPU generations, a HIP-programmed
DCU, and three CPU architectures. No source level change of the
code was required for any of these platforms. The execution space,
the SIMD width, and the ISA-specific load/store specialization are selected
at configure time.

\section{Conclusions}\label{sec:conclusions}

We have described \kwqft, a Kokkos based implementation of Wilson
pure gauge Monte Carlo simulation software for $\suN$ Yang-Mills theory in an
arbitrary number of space-time dimensions. Compile time matrix and lattice
parameters express the $\suN$ and dimensional genericity, while
\texttt{Kokkos::View}, parallel patterns, reductions, and random number
generators decouple the algorithms from backend specific launch and memory
interfaces. A tracked halo layer with
communication/computation overlap provides distributed memory execution,
and on OpenMP and Serial hosts a cross-site SIMD path with ISA-specialized
de-interleaving loads and stores accelerates the same update kernels
without altering the CUDA, HIP, or SYCL per-site device path.

The implementation reproduces the exact two dimensional plaquette for
$\mathrm{SU}(2)$--$\mathrm{SU}(15)$ and published three and
four dimensional values for $N$ up to 17. On an NVIDIA A100 the Kokkos CUDA
backend achieves 81\%--99\% of the throughput of a native CUDA code, with
more than 90\% for $L\geq32$. On SVE-512 hardware the host SIMD path
accelerates the OpenMP build by $1.7$ to $3.5\times$ for $N=2$ to $15$. On the
LineShine LX2 system with MPI+OpenMP, strong scaling of a $128^3\times256$
SU(4) lattice from 16 to 512 nodes reaches a $16.2\times$ speedup, and
trajectory times on six CPU and GPU platforms follow a common $N$
dependence, with a HIP-programmed DCU performing between two NVIDIA GPU
generations. Together, these results support the conclusion that a complete
pure gauge generation and measurement path can be written once in Kokkos
and executed with competitive performance from a single processor to a large
distributed memory supercomputer.

A longer term objective is to extend \kwqft\ from the
pure gauge theory to theories with dynamical fermions, with Wilson-type
Dirac operators whose spin and stencil structures are parameterized by the
space-time dimension in the same way as the gauge kernels. This requires
pseudofermion fields, portable Krylov solvers, gauge and fermion force
kernels, reversible symplectic integrators, and a Hybrid Monte Carlo
algorithm for gauge configuration generation. The gauge field representation,
the halo layer, and the host SIMD infrastructure described here are designed
to be reused by these components.

\section*{Data and code availability}

The \kwqft\ source code is available at
\url{https://github.com/sunwayihep/kwqft} and the results in this paper
can be reproduced by running the code with the parameters described in the paper. 

\printcredits

\section*{Declaration of competing interest}

The author declares that there are no known competing financial interests or
personal relationships that could have appeared to influence the work
reported in this paper.

\section*{Acknowledgements}

The author thanks Ying Chen and Yi-Bo Yang for useful discussions, and
Shuyu Zhang for providing access to the Hygon DCU platform. This work was
supported by the National Key Research and Development Program of China
under Grant No.~2025YFB3003602, the Chinese Academy of Sciences under Grant
No.~YSBR-101, and the National Natural Science Foundation of China under
Grant No.~12205311. The author is also grateful to the Institute of High
Energy Physics (Beijing) and the China Spallation Neutron Source (Dongguan)
for access to their computing resources, and to the National Supercomputing
Center in Shenzhen (NSCC-SZ) for access to the LineShine supercomputer.

\bibliographystyle{model1-num-names}
\bibliography{references}

\end{document}